\documentclass[twocolumn,showpacs,showkeys,preprintnumbers,amsmath,amssymb,twoside]
{revtex4}

\usepackage{color}
\usepackage{graphics}
\usepackage{dcolumn}
\usepackage{bm}
\usepackage{amssymb,amsmath}
\usepackage{latexsym}
\renewcommand\textfraction 0
\renewcommand\topfraction 1
\renewcommand\bottomfraction 1
\begin{document}

\preprint{Submitted to Materials Science and Engineering C
}

\title{Incorporation of cobalt into ZnO nanoclusters}
\author{Igor \surname{Ozerov}}
 \email{ozerov@crmcn.univ-mrs.fr}
\author{Fran\c{c}oise \surname{Chabre}}
\author{Wladimir \surname{Marine}}
 \email{marine@crmcn.univ-mrs.fr}
\affiliation{CRMC-N, UPR 7251 CNRS, Universit\'{e} de la M\'{e}diterran\'{e}e, 
Case 901, 13288 Marseille Cedex 9, France.}
\date{\today} 

\begin{abstract}

The structural, optical and magnetic properties of nanostructured ZnO films
co-doped with cobalt 
and aluminium 
have been studied. The nanocrystalline films, with cluster sizes in range 50 - 100 nm, were 
deposited by pulsed laser ablation in a mixed atmosphere of oxygen and helium. The 
nanocrystallites have the wurtzite structure and are highly oriented with the c-axis 
perpendicular to the substrate. Both optical and electron spin resonance (ESR) spectroscopy 
results show the substitutional incorporation of Co$^{2+}$ ions on the Zn site inside the ZnO 
nanoclusters. The temperature dependence of the ESR spectra follows Curie law 
corresponding to a paramagnetic material. 
\end{abstract}
\keywords{Zinc oxide; Nanoclusters; Cobalt; Laser ablation; Electron spin resonance}
\pacs{76.30.-v, 78.55.Et, 81.05.Ys, 81.15.Fg}

\maketitle


\section{Introduction}

Recently, diluted magnetic semiconductors (DMS) and their nanostructures have attracted a 
lot of 
attention because of their potential applications in magneto-electrical and magneto-optical 
devices \cite{Ohno,Pearton}. Among the other DMS, ZnO is especially interesting because 
the valence and ionic radii of the cations matches those of the magnetic transition metals. This 
fact allows ZnO doping at high concentrations of magnetic ions. Moreover, some recent 
theoretical \cite{Katayama2003} and experimental 
\cite{Ueda2001,Prellier2003,Kim2002,Rode2003} reports have shown the room temperature 
ferromagnetism in ZnO films highly doped with magnetic 
impurities. It is predicted that n-type co-doping should reinforce the ferromagnetic ordering 
\cite{Katayama2003,Ueda2001}.
Two principal approaches are used to introduce the transition metal ions into semiconductor 
materials. First, the magnetitic impurities are implanted into the host matrices 
\cite{Norton2003}. Second, the magnetic elements are introduced during the growth 
\cite{Schwartz}. The second method is preferable for the growth of nanostructures. In low 
dimension nanoscale systems, the surface is widely developed. If the magnetic impurities are 
located at the surface of the nanoclusters, they are subjected to a stochastic crystalline field 
that is hardly controllable. Therefore, growth 
techniques should allow the control of both the size of nanoclusters and the distribution of 
magnetic impurities inside the clusters.

Recently, we have developed a method for synthesizing pure ZnO nanoclusters 
with a low size dispersion by pulsed laser ablation in an atmosphere of 
O$_{2}$ and He \cite{Ozerov2003}. In this paper we extend this method to the synthesis of 
ZnO 
nanoclusters co-doped with cobalt and aluminium. The optical and magnetic properties of the 
nanocrystalline, heavily doped Zn$_{1-x}$Co$_{x}$O:Al films are discussed.

\section{Experimental}

The targets for film deposition were prepared by mixing in appropriate 
proportions of pure ZnO, Co$_{3}$O$_{4}$ and Al$_{2}$O$_{3}$ powders, 
cool pressing in the atmospheric air, and by sintering at a temperature 
of 550$^{\circ}$C for 24 hours. Two series of targets were used for the 
deposition of films having different Al contents (10$^{19}$ and 10$^{20}$ 
cm$^{-3})$. Both series contained targets with three different 
concentrations of Co: 1, 5 and 10 at.{\%}.

The target was placed on a rotating holder inside a stainless steel vacuum 
chamber evacuated by a turbomolecular pump. After pumping the chamber down 
to approximately $2\times 10^{-7}$ mbar, a continuous flux of oxygen mixed with 
helium was introduced into the chamber as an ambient. Differential pumping 
ensured a constant gas flow with controlled partial pressures (4 mbar of 
oxygen and 1.5 mbar of helium) during the film deposition. The silica 
substrate was placed in front of the target and kept at a temperature in range
$380 - 450 ^{\circ}$C. The ablation was performed by a beam of 
pulsed ArF$^{*}$ laser ($\lambda = 193 $ nm, pulse duration 15 ns, FWHM) 
focused onto the target with an incident angle of 45$^{\circ}$ at a fluence of 
3.5 J/cm$^{2}$.

The crystalline structure of the obtained films was analyzed by X-ray 
diffraction (XRD) with Cu K$\alpha$ source. The size and shape of nanoclusters, as 
well as the film morphology were determined by scanning electron 
microscopy (SEM). 

Room temperature optical transmission measurements at near-normal incidence
were performed in the photon energy region from 1.4 to 4.2 eV.
We used a deuterium lamp as the light source and a cooled photo-multiplier in photon 
counting mode linked to a grating monochromator for the detection of the spectra.

The magnetic properties were studied by electron spin resonance (ESR) measurements 
using X-band Bruker EMX spectrometer equipped with an Oxford Instruments helium-flow 
cryostat (4.2 K - 300 K).

\section{Results and Discussion}

\begin{figure}[!ht]
\resizebox{8.5cm}{!}{\includegraphics{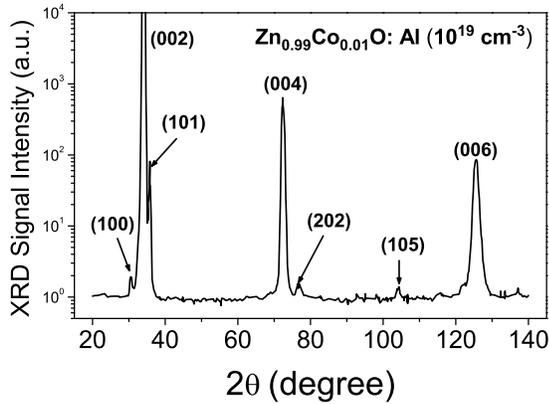}} \caption{\label{XRD}
$\theta $ -- 2$\theta $ X-Ray diffraction pattern 
of a film of Zn$_{0.99}$Co$_{0.01}$O:Al ($10^{19}$cm$^{-3}$). The crystalline planes of 
the wurtzite structure are indicated.}
\end{figure}

\begin{figure}[!hb]
\resizebox{8.5cm}{!}{\includegraphics{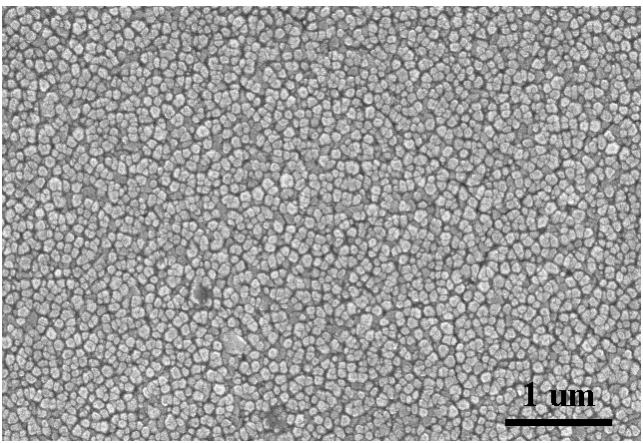}} \caption{\label{SEM}
Typical Scanning Electron Microscopy image of Zn$_{0.95}$Co$_{0.05}$O:Al 
($10^{19}$cm$^{-3}$) film.}
\end{figure}

The XRD results showed that nanostructured films are crystallized in the
wurtzite structure (figure \ref{XRD}). All films are preferentially oriented with the c-axis 
perpendicular to the substrate surface. Even in the case of amorphous substrates, the 
c-axis orientation of the films is favorable energetically because of 
polarity of oxygen- and zinc-terminated (0001) atomic planes of wurtzite 
structure \cite{Ozerov2003}. However, several crystallites have orientations different from 
the preferential one (fig. \ref{XRD}).
Unlike very recent reports \cite{Norton2003,KimKimKim}, no secondary phase highly doped 
with cobalt was detected.

SEM images of various samples show that the film surface morphology weakly depends on 
the deposition temperature in the range from 380 to 450$^{\circ}$C and on the proportion of 
aluminium used. In all cases considered, the distribution of clusters
is homogeneous on the surface (figure \ref{SEM}). The main factor driving a difference 
in the film roughness is the amount of cobalt: the mean size of 
the clusters decreases as the Co content inceases (from 100nm for 1 at. {\%} to 40nm for 10 
at.{\%} of cobalt). Chemical analysis by X-ray fluorescence indicates a good agreement 
between the amount of cobalt used in the targets and the one effectively found in the films.


The obtained films are homogeneous, transparent and have a slight green 
color of ``Rinmans green'' pigment. The intensity of the green color 
increases with cobalt content. From the calibrated deposition rates, we estimated the thickness 
of the films to be approximatively $1 \mu$m \cite{Ozerov2003}.

\begin{figure}[!]
\resizebox{8.5cm}{!}{\includegraphics{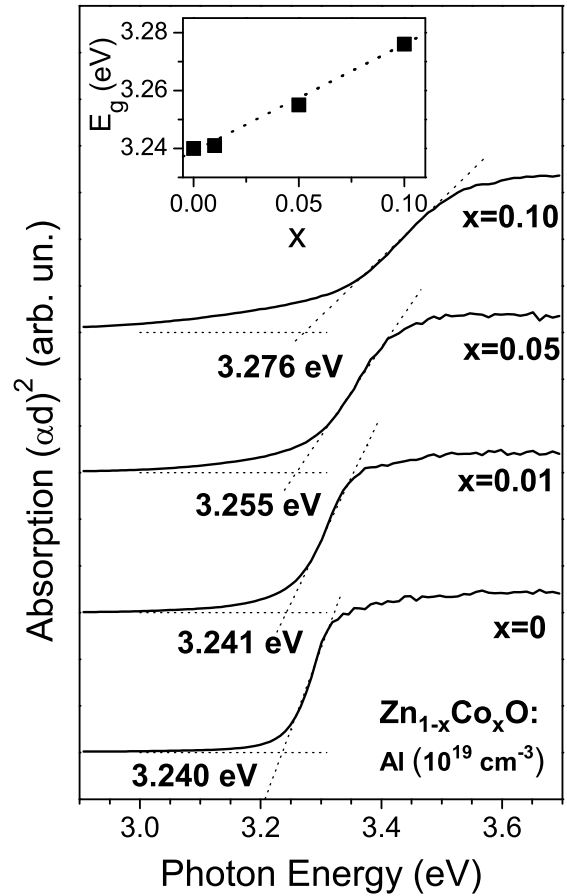}}\caption{\label{figGap}
Optical absorption spectra of Zn$_{1-x}$Co$_{x}$O :Al (10$^{19}$ cm$^{-3}$) films for x 
= 0, 0.01, 0.05 and 0.1. The dotted lines represent linear fits in the band edge region. The inset 
shows the increase of the bandgap with Co concentration.}
\end{figure}

\begin{figure}[!]
\resizebox{8.5cm}{!}{\includegraphics{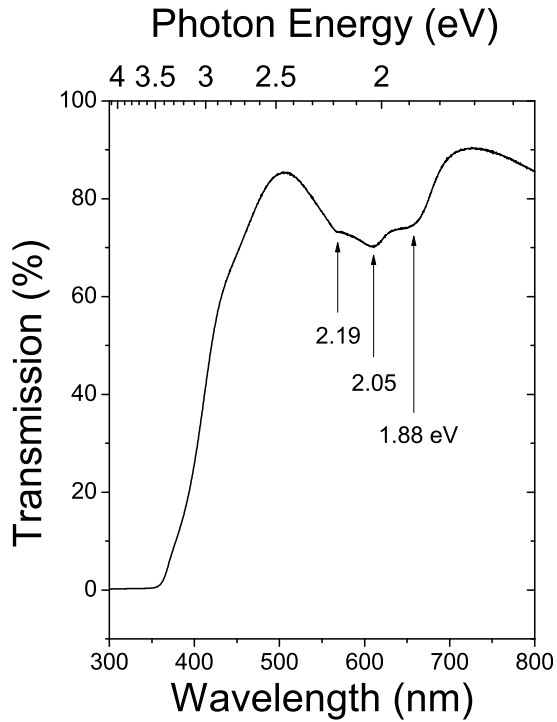}}\caption{\label{figTransm}
Optical transmission spectra of Co-doped ZnO films. The d-d transitions in a Co$^{2+}$ ion 
are indicated with arrows.}
\end{figure}

\begin{figure}[!]
\resizebox{8.5cm}{!}{\includegraphics{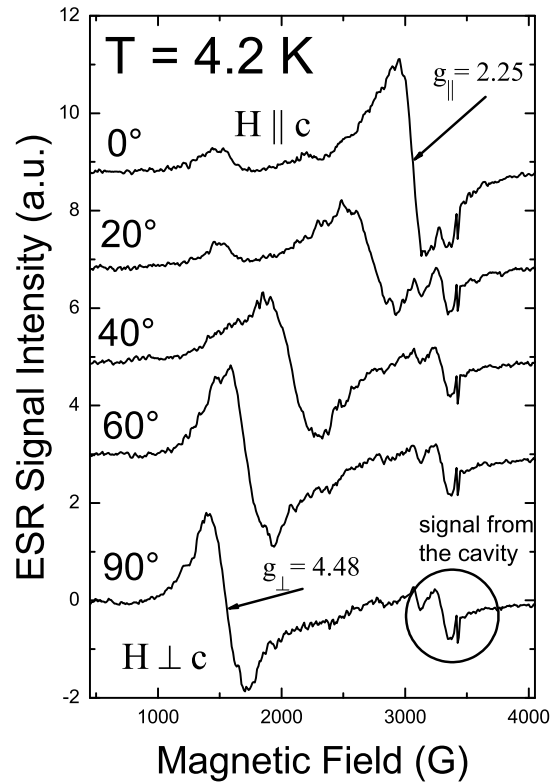}} \caption{\label{FigESRang}
X-band ESR spectra of a Zn$_{0.9}$Co$_{0.1}$O : Al nanocrystalline film. The 
concentration of Al  is 10$^{19}$ cm$^{-3}$. The temperature was 4.2 K.}
\end{figure}

Figure \ref{figGap} shows the absorption spectra of the films at room temperature in the 
photon energy range close to the fundamental absorption edge of ZnO. The absorbance of a
pure nanocrystalline ZnO film is given for comparison.
Zinc oxide is a direct gap semiconductor. Thus for allowed interband transitions 
at photon energies above the band gap $E_{g}$, the absorption 
coefficient depends on the photon energy h$\nu$ as follows: $\alpha(h\nu) \sim \sqrt{h\nu -
E_{g}}$. Consistent with the behavior of a direct gap, the absorption spectra are plotted as 
$(\alpha \cdot $d)$^{2}$, where d is the film thickness. The high-energy parts of the 
absorption spectra were fitted with the linear function (see fig. \ref{figGap}). The 
extrapolation of this line up to zero absorption gives a value of the direct band gap. We found 
that E$_{g}$ increases linearly with increasing Co concentration from 3.240 eV for pure ZnO 
up to 3.276 eV for Zn$_{0.9}$Co$_{0.1}$O (inset in Fig. \ref{figGap}).
These results are in desagreement with the gap narrowing with increase in Co for Zn$_{1-
x}$Co$_{x}$O doped at extremely high concentrations reported in \cite{Kim2002} ($x$ 
varied from 0.08 to 0.22) and in \cite{Lim2003} ($x = 0.1 - 0.4$). The values for E$_{g}$ 
reported in \cite{Kim2002,Lim2003} approach that reported for CoO in \cite{Pratt1959}. On 
the contrary, Yoo et al. \cite{Yoo2001} reported the increase of E$_{g}$ in ZnCoAlO films 
with increase of Co and Al concentrations at moderate doping levels. Thus, the exact nature of 
the gap in the highly Co-doped ZnO films remains uncertain. The absorption is also present 
for the photon energies below the fundamental absorption edge (fig. \ref{figGap}) and it 
becomes stronger with increasing dopant concentration. A strong sp-d exchange interaction 
between the band states and localized d-electrons reported in Co-doped ZnO from magneto-
optical measurements \cite{Ando2001b} is likely responsible for this absorption tail 
expanding below the band gap energy.

The optical transmission in wide spectral range is shown in fig. \ref{figTransm}. Three 
absorption bands situated at the photon energies of 1.88, 2.05 and 2.19 eV (shown by arrows 
in fig. \ref{figTransm}) were identified as d-d transitions in Co$^{2+}$ ions situated in the 
site with a crystalline field
of tetrahedral symmetry \cite{Koidl1977}. These transitions were also observed in highly 
doped ZnCoO films \cite{Kim2002,Lim2003,Yoo2001}. However, these bands are large and 
badly resolved at room temperature, and they can be overlapped with the transitions provided 
by Co in octahedral symmetry
sites \cite{Pratt1959} or by absorption bands in Co$_{3}$O$_{4}$ \cite{Yamamoto2003} 
situated between 1.5 and 2.5 eV. Thus, more local methods like ESR should be used to 
confirm the proper substitutional doping of Co into Zn sites.


Typical ESR spectra of Zn$_{1-x}$Co$_{x}$O nanoclusters are shown in fig. 
\ref{FigESRang}. The measurements have been performed at a temperature of 4.2 K by 
rotating the magnetic field in the plane parallel to the sample surface normal. The spectra 
contain 
only one anisotropic resonance line with a Lorentzian shape.
The lowest resonance field is obtained for the direction of the magnetic field perpendicular to 
the [001] direction of the wurtzite structure, and the highest one corresponds to the parallel 
configuration. Taking \textbf{x} and \textbf{z} axes in the rotation plane, \textbf{z} being 
parallel to [001] direction, the observed resonance transition is described by the following 
effective spin Hamiltonian \cite{Abragam}:

\begin{equation}
\label{effHam}
\mathcal{H}^{eff} = g^{eff}_{\parallel}\beta HS^{eff}_{z} \cos \theta  + g^{eff}_{ \bot }
\beta H S^{eff}_{x} \sin \theta
\end{equation}

\noindent
where $\beta$ is the Bohr magneton, $H$ is the magnetic field,
$g^{eff}_{\parallel}$ and $g^{eff}_{ \bot }$ are the effective values of the g tensor, 
$S^{eff}_{z,x}$ are the projections of the effective spin and $\theta$ is the angle between 
magnetic field and \textbf{z} axis. For the observed transition, taking $S^{eff} = 1/2$, we 
obtain the values $g^{eff}_{\parallel} = 2.25$ and $g^{eff}_{ \bot }= 4.48$ (fig. 
\ref{FigESRang}).

In the ZnO wurtzite structure, the lowest state $^{4}$F of the 3d$^{7}$ Co$^{2 + }$ ion is 
split by the tetrahedral crystal field, $^{4}$A$_{2}$ being the ground state \cite{Koidl1977}. 
The trigonal component of the crystal field splits the ground state $^{4}$A$_{2}$ 
corresponding to spin $S=3/2$ into two Kramers doublets, the lower being S~=~$\pm $1/2, 
and the higher S~=~$\pm $3/2. The following Hamiltonian takes the axial crystalline field 
into account \cite{Abragam,Estle}:

\begin{equation}
\label{Ham}
\mathcal{H} = g_{\parallel} H\beta S_{z} \cos \theta  + g_{\bot } H\beta S_{x} \sin \theta + 
D [ S_{z}^{2} 
- \frac{1}{3} S (S+1) ]
\end{equation}
\noindent
where spin $S = 3/2$ and 2D corresponds to zero-field splitting.  The 
Hamiltonian (\ref{effHam}) can be obtained from (\ref{Ham}) by assuming $D>0$ and 
$\vert D\vert \gg \beta H$. The effective $g$-factor values are: $g^{eff}_{\parallel} = 
g_{\parallel}$ and $g^{eff}_{ \bot } = 2g_{ \bot }(1-\frac{3 (g_{ \bot }\beta 
H)^{2}}{16D^{2}}) \approx 2g_{ \bot }$ \cite{Estle}.
Thus, the only observed transition corresponds to $\vert 
-\raise.5ex\hbox{$\scriptstyle 1$}\kern-.1em/ 
\kern-.15em\lower.25ex\hbox{$\scriptstyle 2$}\rangle\rightarrow~\vert 
+\raise.5ex\hbox{$\scriptstyle 1$}\kern-.1em/ 
\kern-.15em\lower.25ex\hbox{$\scriptstyle 2$}\rangle$ absorption in the lower Kramers 
doublet.
The transitions between doublets can not be achieved in the available magnetic field range 
because of the large zero-field splitting 2D \cite{Estle,Jedercy2004,Isber1995}.

The angular dependence of the spectra of the Co$^{2+}$ ions confirms the hypothesis of 
substitutional incorporation of dopant in zinc sites. No signal with other symmetry was 
detected.
In the spectra of several samples a low intensity peak at 1555~G was observed for $H 
\parallel c$ (fig. \ref{FigESRang}). This transition is due to minority nanocrystallites with the 
orientations different from that of the main part of the film which is oriented with the c-axis 
perpendicular to the substrate. This observation is in agreement with the results of XRD 
experiments (fig. \ref{XRD}).

The width $\Delta H _{pp}$ of the Co$^{2+}$ ESR resonance line is independent on the 
temperature and varies with cobalt concentration from 250 G for Zn$_{0.99}$Co$_{0.01}$O 
to 360 G for Zn$_{0.90}$Co$_{0.10}$O in the perpendicular configuration $H \bot c.$ 
These line width values can not be explained by a non-resolved hyperfine structure from 
$^{59}$Co (nuclear spin 7/2, 100{\%}), the components of the hyperfine tensor being too 
small $A_{\parallel} = 16.11\times 10^{-4}$ cm$^{-1}$ an A$_{\bot} = 3.00\times 10^{-
4}$ cm$^{-1}$ \cite{Estle,Jedercy2004}. The line width increased due to dipole-dipole 
interactions. We estimate the dipole-dipole broadening from $W_{dip-dip}= -(\mu H_{dip}) 
\approx (g \beta S)^2 N$, where spin $S=3/2$ and $N$ is cobalt concentration. $W_{dip-
dip}$ is larger than 200 Gauss for typical concentrations of cobalt we used, that is consistent 
with the observed line width. No expected exchange narrowing was observed with an increase 
in concentration. 


If the line width does not vary with the temperature, the spin susceptibility $\chi$ is reciprocal 
to the amplitude of the ESR signal \cite{Abragam}. For all the films we deposited, $\chi$ 
follows a Curie law $\chi \propto $ 1/T, where T is the temperature
. This behavior shows that the samples are paramagnetic, meaning that no high-temperature 
ferromagnetism is observed in the Co-doped ZnO nanoclusters. 

\section{Conclusions}

We have developed a method of synthesis of high quality nanostructured ZnO films doped 
with Co and Al. The sizes and crystalline quality are controlled by the laser parameters 
(fluence) and by the ambient gas pressure. The films are constituted by 
nanocrystallites with wurtzite structure, oriented with the c-axis 
perpendicular to the substrate. The optical and radiospectroscopic measurements show the 
homogenous distribution of the cobalt ions in substitution of zinc ions inside the nanoclusters.
The magnetic susceptibility follows the Curie law characteristic for a paramagnetic material.

\begin{acknowledgments} 
It is a pleasure to thank S.~Nitsche for the SEM measurements, as well as V.~Safarov, 
J.~Marfaing and A.~Stepanov for useful discussions.
\end{acknowledgments}


\begin{thebibliography}{99}

\bibitem{Ohno}H. Ohno, Science \textbf{281} (1998) 951.

\bibitem{Pearton} S.J. Pearton, C.R. Abernathy, M.E. Overberg, G.T. Thaler, D. P. Norton, 
N. Theodoropoulou, A.F. Hebard, Y.D. Park, F. Ren, J. Kim, L. A. Boatner, J. Appl. Phys. 
\textbf{93}, 1 (2003).

\bibitem{Katayama2003}H. Katayama-Yoshida, K. Sato, Physica B \textbf{327}, 337 
(2003).

\bibitem{Ueda2001}K. Ueda, H. Tabata, T. Kawai, Appl. Phys. Lett. \textbf{79}, 988 (2001).

\bibitem{Prellier2003}W. Prellier, A. Fouchet, B. Mercey, Ch. Simon, B. Raveau, Appl. 
Phys. Lett. 
\textbf{82}, 3490 (2003).

\bibitem{Kim2002}K.J. Kim, Y.R. Park, Appl. Phys. Lett. \textbf{81}, 1420 (2002).

\bibitem{Rode2003}K. Rode, A. Anane, R. Mattana, J.-P. Contour, O. Durand, R. 
LeBourgeois, J. 
Appl. Phys. \textbf{93}, 7676 (2003).

\bibitem{Norton2003}D.P. Norton, M.E. Overberg, S.J. Pearton, K. Pruessner, J.D. Budai , 
L.A. Boatner, M.F. Chisholm, J.S. Lee, Z.G. Khim, Y.D. Park, R.G. Wilson, Appl. Phys. Lett. 
\textbf{83}, 5488 (2004).

\bibitem{Schwartz}D.A. Schwartz, N.S. Norberg, Q.P. Nguyen, J.M. Parker, and D.R. 
Gamelin,
J. Am. Chem. Soc. \textbf{125}, 13205 (2003).

\bibitem{Ozerov2003}I. Ozerov, D. Nelson, A.V. Bulgakov, W. Marine, M. Sentis, Appl. 
Surf. Sci. 
\textbf{212-213}, 349 (2003).

\bibitem{KimKimKim}J.H. Kim, H. Kim, D. Kim, Y.E. Ihm, W.K. Choo, J. Appl. Phys. 
\textbf{92}, 6066 (2002).

\bibitem{Lim2003}S.W. Lim, D.K. Hwang, J.M. Myoung, Solid State Comm. \textbf{125}, 
231 (2003).

\bibitem{Pratt1959}G.W. Pratt Jr., R. Coelho, Phys. Rev \textbf{116}, 281 (1959).

\bibitem{Yoo2001}Y.Z. Yoo, T. Fukumura, Z. Jin, K. Hasegawa, M. Kawasaki, P. Ahmet, T. 
Chikyow, H. Koniuma, J. Appl. Phys. \textbf{90} 4246 (2001).

\bibitem{Ando2001b}K.Ando, H.Saito, Z. Jin, T.Fukumura, M. Kawasaki, Y. Matsumoto, H. 
Koinuma, J. Appl. Phys. \textbf{89}, 7284 (2001).

\bibitem{Koidl1977}P. Koidl, Phys. Rev. B \textbf{15}, 2493 (1977).

\bibitem{Yamamoto2003}H. Yamamoto, S. Tanaka, K. Hirao, J. Appl. Phys. \textbf{93} 
4158 (2003).

\bibitem{Abragam} A. Abragam, B. Bleaney, Electron Paramagnetic Resonance of Transition 
Ions, Vol. 1, Clarendon Press, Oxford, 1970.

\bibitem{Estle} M. De Wit and T.L. Estle, Bull. Am. Phys. Soc. \textbf{6}, 445 (1961).

\bibitem{Jedercy2004} N. Jedercy, H.J. van Bardeleben, Y. Zheng, J.L. Cantin, Phys. Rev. B 
\textbf{69}, 041308(R) (2004).

\bibitem{Isber1995}S. Isber, M. Averous, Y. Shapira, V. Bindilatti, A.N. Anisimov, N.F. 
Oliveira Jr, V.M. Orera, M. Demianiuk, Phys. Rev. B \textbf{51}, 15211 
(1995).
\end{thebibliography}
\end{document}